\documentclass[fleqn,twoside]{article}
\usepackage{espcrc2,epsfig}

\usepackage{graphicx}
\usepackage[figuresright]{rotating}

\newcommand{\nn}{\nonumber}
\newcommand{\beq} {\begin{equation}}
\newcommand{\eeq} {\end{equation}}
\newcommand{\beqa} {\begin{eqnarray}}
\newcommand{\eeqa} {\end{eqnarray}}

\newcommand{\ie}{{\it i.e.}}
\newcommand{\eg}{{\it e.g.}}

\newcommand{\order}[1]{${\cal O}\left(#1 \right)$}

\newcommand{\eq}[1]{(\ref{#1})}

\newcommand{\qu}{{\rm q}}

\newcommand{\qbm}{{\rm\bar q}}

\title{\vskip -30pt %
{\hbox to\hsize{\normalsize\hfil\rm  26 September, 2002}} \vskip -3pt
{\hbox to\hsize{\normalsize\hfil\rm HIP-2002-43/TH}}
\vskip 5pt
Structure Functions are not Parton Probabilities\thanks{Talk at ICHEP 2002, Amsterdam (July 2002).}}

\author{Paul Hoyer\address{Department of Physical Sciences and Helsinki Institute of Physics\\ 
        P.O. Box 64, FIN-00014 Helsinki University, Finland}%
        \thanks{Address until 1 August 2002: Nordita, Blegdamsvej 17, DK-2100 Copenhagen, Denmark. Research supported in part by the
European Commission under contract HPRN-CT-2000-00130.}}
       
\begin{document}

\begin{abstract}
Parton distributions given by deep inelastic lepton scattering (DIS) are not equal to the probabilities of finding those partons in the parent wave function. Soft rescattering of the struck parton within the coherence length of the hard process influences the DIS cross section and gives dynamical phases to the scattering amplitudes. This gives rise to diffractive DIS, shadowing in nuclear targets and transverse spin asymmetry.
\vspace{1pc}
\end{abstract}

\maketitle

\section{DIS DYNAMICS}

Following the pioneering work of Drell, Levy and Yan \cite{Drell:1969km} in 1969, the deep inelastic scattering ($\ell N \to \ell + X$, DIS) cross section has been commonly thought to be given by the probability of finding a parton in the target $N$ wave function, \ie, by the parton density. In the Bjorken limit of photon energy $\nu$ and virtuality $Q^2$ tending to infinity at fixed $x_B=Q^2/2m_N \nu$, final state interactions do not affect the DIS cross section -- in a scalar theory such as considered in Ref. \cite{Drell:1969km}. In QCD, gluon exchange modifies the relation between the cross section and the parton density \cite{Brodsky:2002ue}, and causes observed effects such as diffraction, shadowing and transverse spin asymmetry. In this talk I give a brief account of this.

%%%%%%%%%%%%%%%%%%%%%%%%%
\begin{figure}[hbt]
\begin{center}
\leavevmode
{\epsfxsize=7truecm \epsfbox{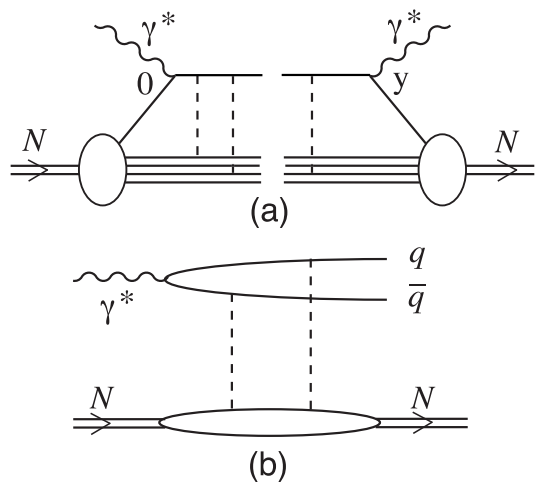}}
\end{center} \vspace{-1cm}
\caption[*]{(a) The DIS cross section and (b) a diffractive DIS amplitude. The photon moves along the negative $z$-axis in (a) and in the opposite direction in (b).} \vspace{-.5cm}
\label{fig1dis}
\end{figure}
%%%%%%%%%%%%%%%%%%%%%%%%%

The standard DIS `handbag' shown in Fig. 1a may be viewed as a Light-Cone (LC) time-ordered diagram where the ordering of the vertices from left to right corresponds to an ordering in LC time $y^+ = y^0+y^3$. In a frame where the virtual photon moves along the negative $z$-axis $(q^- \simeq 2\nu,\ q^+ \simeq -m_N x_B)$ there are no $\gamma^* \to q\bar q$ splittings of the kind shown in Fig. 1b, since the quark lines must have $p^+ > 0$. Due to the large value of $q^-$ the scattering takes place almost instantaneously in LC time, $y^+ \sim 1/\nu$.

The dashed lines in Fig. 1a indicate that the struck quark can (and typically does) rescatter elastically on the color field of the target. Since the exchanged gluon has spin 1 this soft (instantaneous, coulombic) scattering does not decrease with the energy $\nu$ of the quark -- whereas scalar exchange would bring a factor $\propto 1/\nu$. Thus rescattering affects the DIS cross section at leading twist if it occurs within the coherence length $\nu/Q^2 = 1/2m_Nx_B$ of the virtual photon. Even at moderately small values of $x_B$ the coherence length is comparable to the size (several fm) of a nuclear target. Note that the rescattering occurs in vanishing LC time $y^+$ due to the relativistic motion of the struck quark. The DIS amplitudes nevertheless acquire complex dynamical phases from contributions of on-shell intermediate states between the rescatterings. The interference between amplitudes with and without rescattering gives rise to diffraction and shadowing \cite{Piller:2000wx}.

\section{SUBTLETIES OF LC GAUGE}

QCD factorization in DIS determines the expression for the parton distribution in a general gauge to be \cite{Collins:1981uw}
\beqa \label{melm}
f_{\qu/N}(x_B,Q^2)= \frac{1}{8\pi} \int dy^- \exp(-ix_B p^+ y^-/2)
\nonumber&&\\
\times
\langle N(p)| \qbm(y^-) \gamma^+\,\big[y^-;0\big] \qu(0)|N(p)\rangle && 
\eeqa
where all fields are evaluated at $y^+=y_\perp =0$. The path-ordered exponential 
\beq \label{poe}
\big[y^-;0\big] \equiv {\rm P}\exp\left[ig\int_0^{y^-}
dw^- A^+(w^-) \right]
\eeq
ensures the gauge invariance of the matrix element and contains the rescattering effects discussed above.

The exponential \eq{poe} reduces to unity in LC gauge ($A^+ = 0$). The matrix element \eq{melm} can then be expressed as an overlap of the target Fock state amplitudes $\psi_n$ at $y^+=0$ \cite{spd},
\beqa \label{prob}
f_{\qu/N}\ {\buildrel {A^+ \to 0} \over \longrightarrow}\ {\cal P}_{\qu/N} = \sum_n
\int \prod_i\, dx_i\, d^2k_{\perp
i} && \nn\\ \times
|\psi_n(x_i,k_{\perp i})|^2 \sum_{j=q} \delta(x_B-x_j) && 
\eeqa 
This appears like the probability of finding a quark in the initial state and is the origin of the misconception that the DIS cross section is an incoherent sum over the target quark density. However, the rescattering and interference effects discussed above have not vanished, they are just hidden by the peculiarities of LC gauge.

In Ref. \cite{Brodsky:2002ue} the properties of a perturbative, abelian model of DIS was studied in Feynman and LC gauge. In the $x_B \to 0$ limit the dominant amplitudes were found to have {\em on-shell} intermediate states between the Coulomb gluon exchanges (dashed lines in Fig. 1). Such states can kinematically occur only {\em after} the $\gamma^*$ has been absorbed, and thus belong to the final state. Each on-shell rescattering amplitude is gauge invariant - hence the on-shell contributions cannot be eliminated by a gauge choice.

The model furthermore showed that in Feynman gauge the DIS cross section is affected only by diagrams where the struck quark rescatters. Indeed, final state interactions (FSI) between spectators cannot affect the DIS cross section at leading twist in this gauge (see, \eg, \cite{Brodsky:2002ue} for a proof). 

In LC gauge the situation was found to be reversed. Rescattering of the {\em struck quark} does not affect the DIS cross section, which is consistent with the fact that the path-ordered exponential $\eq{poe}$ reduces to unity for $A^+=0$. Hence the reinteractions producing on-shell intermediate states in LC gauge {\em occur between spectators}. The singularity of the gluon propagator ${\cal D}_{LC}^{\mu\nu}(k)$ at $k^+=0$ in LC gauge enhances the $k^+ \simeq 0$ exchanges between spectators so that they modify the DIS cross section at leading twist \cite{sp}. 

The spectator interactions occur within a vanishing LC time $y^+$ of \order{1/\nu} after the $\gamma^*$ interacts, even though they are spread over a longitudinal distance of ${\cal O}$(fm) in the target. Hence they can {\em formally} be included in the target Fock amplitudes $\psi_n(y^+=0)$. The $k^+ = 0$ gluon exchanges are then `zero modes' which propagate along the light-like $(y^+=0)$ quantization surface of the Fock states. The Fock state wave functions thus defined are complex, with the rescattering and interference effects contained within their norm. However, the inclusion of the rescattering effects as zero modes of the target LC wave function appears artificial from a physical point of view.

\section{REMARKS}

The insight that DIS dynamics involves rescattering and interference effects has led to interesting further developments. Brodsky, Hwang and Schmidt \cite{Brodsky:2002cx} showed that the complex phase arising from rescattering causes a transverse spin asymmetry at leading twist in semi-inclusive DIS. Thus the azimuthal distribution of the current jet around the virtual photon momentum depends on the transverse polarization of the target nucleon in $\gamma^* + N_\uparrow \to jet+ X$. This ``Sivers effect'' \cite{siv} had previously been thought to be of higher twist \cite{jcc}. The effect is unrelated to the transverse spin carried by the target quark and thus complicates the measurement of the quark transversity distribution. 

In semi-inclusive DIS the quark fields of the parton distribution \eq{melm} are at non-zero transverse separation, $y_\perp \neq 0$. The integral in the path-ordered exponential \eq{poe} must then cover also this transverse distance (at asymptotic LC time) \cite{Ji:2002aa}. The transverse path involves $A^\perp$ and does not vanish in LC gauge -- in fact it is the origin of the transverse spin asymmetry.

Diffractive DIS is described by two-gluon (Pomeron) exchange amplitudes like the one in Fig. 1b. The scattering is most naturally viewed in a frame where the virtual photon moves along the positive $z$-axis, \ie, with $q^+ \simeq 2\nu$. In this frame there is a $\gamma^* \to q\bar q$ transition which at low $x_B$ typically occurs long before the target. Crossing symmetry requires the amplitude to be dominantly imaginary at high energy. 

The frame of Fig. 1b is related to that of Fig. 1a (where $q^- \simeq 2\nu$) by a rotation of $180^\circ$ around the $y$-axis. Due to the nonperturbative nature of the target this rotation cannot in general be performed. The amplitude of Fig. 1b is often interpreted as corresponding, in the frame of Fig. 1a, to scattering off a `Pomeron component' in the target wave function. The dynamics in the two frames can, however, be unambiguously related within the model of Ref. \cite{Brodsky:2002ue}. There are no miracles -- the amplitude remains imaginary after the rotation, and this phase arises from a rescattering in the final state. Thus views of the Pomeron as residing in the nucleon wave function before the $\gamma^*$ interacts are incorrect. 

The above developments motivate comparisons of DIS with other hard processes and renewed checks of the universality of parton distributions. It has become clear that the transverse spin asymmetry has opposite sign in DIS and the Drell-Yan process \cite{jcc,Brodsky:2002rv}. Furthermore, the dynamics of the rescattering which gives rise to shadowing is quite different in these two processes \cite{Peigne:2002iw}.

\vspace{.3cm}
{\bf Acknowledgments.} The work I have described was done in collaboration with S. J. Brodsky, N. Marchal, S. Peign\'e and F. Sannino. I am also grateful for helpful discussions with M. Diehl and X. Ji.


\vspace{.5cm}

\begin{thebibliography}{9}


%\cite{Drell:1969km}
\bibitem{Drell:1969km}
S.~D.~Drell and T.~M.~Yan,
%``Connection Of Elastic Electromagnetic Nucleon Form-Factors At Large Q**2 And Deep Inelastic Structure Functions Near Threshold,''
Phys.\ Rev.\ Lett.\  {\bf 24} (1970) 181;\\
%%CITATION = PRLTA,24,181;%%
%
%\cite{Drell:1969wd}
%\bibitem{Drell:1969wd}
S.~D.~Drell, D.~J.~Levy and T.~M.~Yan,
%``A Theory Of Deep Inelastic Lepton Nucleon Scattering And Lepton Pair Annihilation Processes. 2. Deep Inelastic Electron Scattering,''
Phys.\ Rev.\ D {\bf 1} (1970) 1035.
%%CITATION = PHRVA,D1,1035;%%

%\cite{Brodsky:2002ue}
\bibitem{Brodsky:2002ue}
S.~J.~Brodsky, P.~Hoyer, N.~Marchal, S.~Peigne and F.~Sannino,
%``Structure functions are not parton probabilities,''
Phys.\ Rev.\ D {\bf 65} (2002) 114025
[arXiv:hep-ph/0104291].
%%CITATION = HEP-PH 0104291;%%

%\cite{Piller:2000wx}
\bibitem{Piller:2000wx}
G.~Piller and W.~Weise,
%``Nuclear deep-inelastic lepton scattering and coherence phenomena,''
Phys.\ Rept.\  {\bf 330}, (2000) 1 [arXiv:hep-ph/9908230].
%%CITATION = HEP-PH 9908230;%%

%\cite{Collins:1981uw}
\bibitem{Collins:1981uw}
J.~C.~Collins and D.~E.~Soper,
%``Parton Distribution And Decay Functions,''
Nucl.\ Phys.\ B {\bf 194} (1982) 445.
%%CITATION = NUPHA,B194,445;%%

\bibitem{spd}
S. J. Brodsky, M. Diehl and D. S. Hwang, Nucl.\ Phys.\ B {\bf 596} (2001) 99
[arXiv:hep-ph/0009254]; \\
M. Diehl, T. Feldmann, R. Jakob and P. Kroll, Nucl.\ Phys.\ B {\bf 596} (2001) 33 [arXiv:hep-ph/0009255].

\bibitem{sp} A. Metz and S. Peign\'e, Private communication.

\bibitem{sjb} S. J. Brodsky, Private communication.

%\cite{Brodsky:2002cx}
\bibitem{Brodsky:2002cx}
S.~J.~Brodsky, D.~S.~Hwang and I.~Schmidt,
%``Final-state interactions and single-spin asymmetries in semi-inclusive  deep inelastic scattering,''
Phys.\ Lett.\ B {\bf 530} (2002) 99
[arXiv:hep-ph/0201296].
%%CITATION = HEP-PH 0201296;%%

\bibitem{siv} D. W. Sivers, Phys.\ Rev.\ D {\bf 41} (1990) 83 and Phys.\ Rev.\ D {\bf 43} (1991) 261.

\bibitem{jcc} J. C. Collins, Nucl.\ Phys.\ B {\bf 396} (1993) 161 [arXiv:9208213] and
%
%\cite{Collins:2002kn}
%\bibitem{Collins:2002kn}
%J.~C.~Collins,
%``Leading-twist single-transverse-spin asymmetries: Drell-Yan and  deep-inelastic scattering,''
Phys.\ Lett.\ B {\bf 536} (2002) 43
[arXiv:hep-ph/0204004].
%%CITATION = HEP-PH 0204004;%%

%\cite{Ji:2002aa}
\bibitem{Ji:2002aa}
X.~Ji and F.~Yuan,
%``Parton distributions in light-cone gauge: Where are the final-state  interactions?,''
Phys.\ Lett.\ B {\bf 543} (2002) 66
[arXiv:hep-ph/0206057]; \\
%%CITATION = HEP-PH 0206057;%%
%
%\cite{Belitsky:2002sm}
%\bibitem{Belitsky:2002sm}
A.~V.~Belitsky, X.~Ji and F.~Yuan,
%``Final state interactions and gauge invariant parton distributions,''
arXiv:hep-ph/0208038.
%%CITATION = HEP-PH 0208038;%%

%\cite{Brodsky:2002rv}
\bibitem{Brodsky:2002rv}
S.~J.~Brodsky, D.~S.~Hwang and I.~Schmidt,
%``Initial-state interactions and single-spin asymmetries in Drell-Yan  processes,''
arXiv:hep-ph/0206259.
%%CITATION = HEP-PH 0206259;%%

%\cite{Peigne:2002iw}
\bibitem{Peigne:2002iw}
S.~Peigne,
%``Absence of shadowing in Drell-Yan production at finite transverse  momentum exchange,''
arXiv:hep-ph/0206138.
%%CITATION = HEP-PH 0206138;%%


\end{thebibliography}
\end{document}